\documentclass[preprint,12pt,english]{elsarticle}

\usepackage{graphicx}
\usepackage{subfigure}
\usepackage{amssymb}
\usepackage{amsmath}
\usepackage{lineno}
\usepackage{color}
\usepackage[section]{placeins}
\usepackage[T1]{fontenc}
\usepackage{float}

\pdfoutput=1

\begin{document}

\begin{frontmatter}

%% Title, authors and addresses

%% use the tnoteref command within \title for footnotes;
%% use the tnotetext command for the associated footnote;
%% use the fnref command within \author or \address for footnotes;
%% use the fntext command for the associated footnote;
%% use the corref command within \author for corresponding author footnotes;
%% use the cortext command for the associated footnote;
%% use the ead command for the email address,
%% and the form \ead[url] for the home page:
%%
%% \title{Title\tnoteref{label1}}
%% \tnotetext[label1]{}
%% \author{Name\corref{cor1}\fnref{label2}}
%% \ead{email address}
%% \ead[url]{home page}
%% \fntext[label2]{}
%% \cortext[cor1]{}
%% \address{Address\fnref{label3}}
%% \fntext[label3]{}

\title{Neutron Calibration Sources in the Daya Bay Experiment}

%% use optional labels to link authors explicitly to addresses:
%% \author[label1,label2]{<author name>}
%% \address[label1]{<address>}
%% \address[label2]{<address>}

\author[2,1]{J.~Liu\corref{cor1}}
\author[1]{R.~Carr}
%\author[1]{R.~Cortez\fnref{fn1}}
\author[1,4]{D.~A.~Dwyer}
\author[2]{W.~Q.~Gu}
\author[2]{G.~S.~Li\corref{cor2}}
%\author[1]{J.~Pendlay}
\author[1,3]{R.~D.~McKeown}
\author[1,5]{X.~Qian}
\author[1,6]{R.~H.~M.~Tsang}
\author[1]{F.~F.~Wu}
\author[1,5]{C.~Zhang}

\address[1]{Kellogg Radiation Laboratory, California Institute of Technology, 
Pasadena, California, USA}
\address[2]{Department of Physics, Shanghai Jiao Tong University, Shanghai, 
China}
\address[3]{Department of Physics, College of William and Mary, Williamsburg, Virginia, USA}
\address[4]{Lawrence Berkeley National Laboratory, Berkeley, California, USA}
\address[5]{Brookhaven National Laboratory, Upton, New York, USA}
\address[6]{Department of Physics and Astronomy, University of Alabama, Tuscaloosa, Alabama 35487, USA}

\cortext[cor2]{lgs1029@sjtu.edu.cn}
\cortext[cor1]{jianglai.liu@sjtu.edu.cn}
%%\fntext[fn1]{Deceased}

\begin{abstract}
%% Text of abstract
We describe the design and construction of the 
low rate neutron calibration sources used in the Daya Bay Reactor 
Anti-neutrino Experiment. 
Such sources are free of correlated gamma-neutron emission, which is 
essential in minimizing induced background
in the anti-neutrino detector. The design characteristics have been validated
in the Daya Bay anti-neutrino detector.
\end{abstract}

\begin{keyword}
%% keywords here, in the form: keyword \sep keyword

neutron sources \sep $^{241}$Am-$^{13}$C \sep reactor neutrinos \sep $\theta_{13}$ \sep Daya Bay \sep calibration

%% MSC codes here, in the form: \MSC code \sep code
%% or \MSC[2008] code \sep code (2000 is the default)

\end{keyword}

\end{frontmatter}

%%
%% Start line numbering here if you want
%%
%%\linenumbers

%% main text
\section{Introduction}
\label{sec:intro}
Neutron sources are important calibration sources with a wide 
range of applications. 
%We describe a source specifically 
%designed for the Daya Bay Reactor Anti-neutrino Experiment. 
In modern reactor neutrino experiments such as Daya Bay \cite{tdr},
Double Chooze \cite{DCTDR} and RENO \cite{RENOTDR},
the electron anti-neutrinos are detected by liquid scintillator 
detectors via the inverse beta decay (IBD) reaction 
$\bar{\nu_e} + p \rightarrow e^+ + n$ with the time-correlated 
prompt positron signal ranging from 1 to 10 MeV, and the delayed neutron capture signal 
of $\sim$8 MeV on the gadolinium dopant or 2.2 MeV on hydrogen.
In the Daya Bay experiment, which is located in South China, 
the regular deployment of neutron sources 
allows a thorough characterization of the detector response 
to IBD neutrons, which contributes to the recent discovery of the 
neutrino mixing angle $\theta_{13}$~\cite{DYB12}. 

The automated calibration units (ACUs) of the Daya Bay anti-neutrino 
detectors (ADs) are detailed in~\cite{ref:ACU}. 
Each AD is submerged in a water pool and is equipped with three ACUs 
on top. Neutrino interactions are rare, so minimizing potential 
background created by the neutron 
sources is a top consideration. In this article, we discuss the 
design and construction of low rate ($\sim<$ 1 Hz) 
neutron sources that are free of correlated gamma emission. 
Although used in a specific experiment, such a source 
could also be potentially useful in other occasions 
where ultra low background is desired, or where one seeks a 
neutron source with no associated gamma rays.

\section{Physical requirements to the neutron sources}
\label{sec:requirements}
As discussed in \cite{DYB12}, the Daya Bay ADs
are arranged in a 4-near and 4-far configuration to the nuclear reactor 
cores, with 
the near detectors sampling the reactor neutrino flux and far detectors 
detecting the $\bar{\nu_e}$ disappearance due to $\theta_{13}$. The rates 
of the IBD in 
the near (far) detectors are approximately 700 (70)/day/AD~\cite{DYB12}. 
For each AD module, 
three ACUs are instrumented, each of which is capable of deploying radioactive 
sources vertically 
into the detector. There is one neutron 
source in each ACU~\cite{tdr}\cite{ref:ACU}.
During normal neutrino data taking, the neutron sources are ``parked'' inside
the ACUs right above the AD. Although these neutrons rarely leak directly 
into the neutrino target (gadolinium loaded liquid scintillator),
they can get captured on surrounding materials, in particular, stainless steel, 
i.e. Fe, Mn, Cr, Ni, etc, and emit gamma rays 
ranging from 6 to 10 MeV. These high energy gamma rays, 
later referred to as SS-capture gammas, are difficult 
to shield, leading to two kinds of background due to accidental 
and correlated coincidences: a) the SS-capture gammas
can be in random delayed coincidence with ambient gamma
background, mimicking the IBD signals, and b) if multiple neutrons 
are emitted per decay, or a neutron is emitted with a correlated gamma ray, 
or a neutron is producing gammas via inelastic scattering, they 
can form real correlated background to the IBDs. The 
signal to background ratio at the far site 
drives the requirements to the neutron sources:
\begin{itemize}
\item the accidental background to be less than 
5\% of the IBD signal at the far site. Such 
a background can be statistically subtracted;
\item the correlated background to be less than 0.5\% of the 
IBD signal at the far site, i.e. $<0.35$ per day. Such a background can 
be estimated via 
Monte Carlo (MC) simulation and benchmarked by special control data. 
\end{itemize}

For a typical neutron source inside the ACU, a GEANT4 \cite{G4} simulation with 
realistic detector geometry predicts that the SS-capture gamma ray leaking 
into the detector satisfying the IBD delay energy cut is approximately 
$2\times10^{-3}$ per neutron. Taking into account the 
$\sim$70 Hz singles rate~\cite{DYB12} and 200 $\mu$s coincidence window, 
the first requirement translates into a limit of the neutron rate per source
$<$1 Hz.

\section{Design of the neutron source}
\label{sec:design}

\subsection{Selection of neutron source}
There are several types of commonly used compact neutron 
calibration sources, e.g. fission sources 
such as $^{252}$Cf, ($\alpha$,n) sources such as $^{241}$Am-Be, 
and photo-neutron sources such as $^{124}$Sb-Be
. For the third type, photo-neutron cross section is of the order millibarns, 
implying the need for a rather strong driving gamma source. Due to 
low background considerations, this option was rejected early on. 

A $^{252}$Cf source emits multiple neutrons per fission together with gammas. 
Typical ($\alpha$,n) sources have correlated gamma neutron emission 
when the final state nucleus is in an 
excited state. As mentioned earlier, 
such sources would inevitably lead to correlated background in the AD, 
in addition to the accidental 
background. Just to set the scale, the predicted correlated 
background for $^{252}$Cf and $^{241}$Am-Be sources are 2.6/day and 1.3/day, 
respectively, assuming a 0.5 Hz neutron rate.

The alphas from $^{241}$Am are $\sim$5.5 MeV. 
To eliminate correlated gamma rays emission, 
$^{7}$Li would be a good candidate target since such alpha energy
can only produce ground state $^{10}$B.
However, the most common and chemically inert Li compound is LiF, and 
($\alpha$,n) on $^{19}$F creates a significant amount of high energy gamma rays.

$^{13}$C is the final candidate. We note that $^{241}$Am-$^{13}$C can produce 
neutrons with the final state $^{16}$O either in the ground state, and 
the first or second excited state. The first excited state of $^{16}$O decays 
into a e$^+$e$^-$ 
pair, which will be stopped by the source enclosure and surrounding materials
\footnote{The stopped e$^+$ would annihilate into two back-to-back 0.511 MeV gammas, 
which can hardly deposit enough energy in the AD to cross the trigger threshold.}.
The second excited state of $^{16}$O will emit a 6.13 MeV gamma ray, 
producing correlated background together with the neutron. However if the energy of $\alpha$ is 
attenuated 
to below 5.11 MeV, this correlated $\gamma$-neutron process can be eliminated 
entirely. Based on all considerations above, we selected
$^{241}$Am-$^{13}$C as the neutron source, further requiring that 
$E_{\alpha} < 5.11$ MeV.

\section{Physical Design}
\subsection{$^{241}$Am Sources}
\label{sec:alpha}
\begin{figure}[!htbp]
  \centering
  \includegraphics[width=4in]{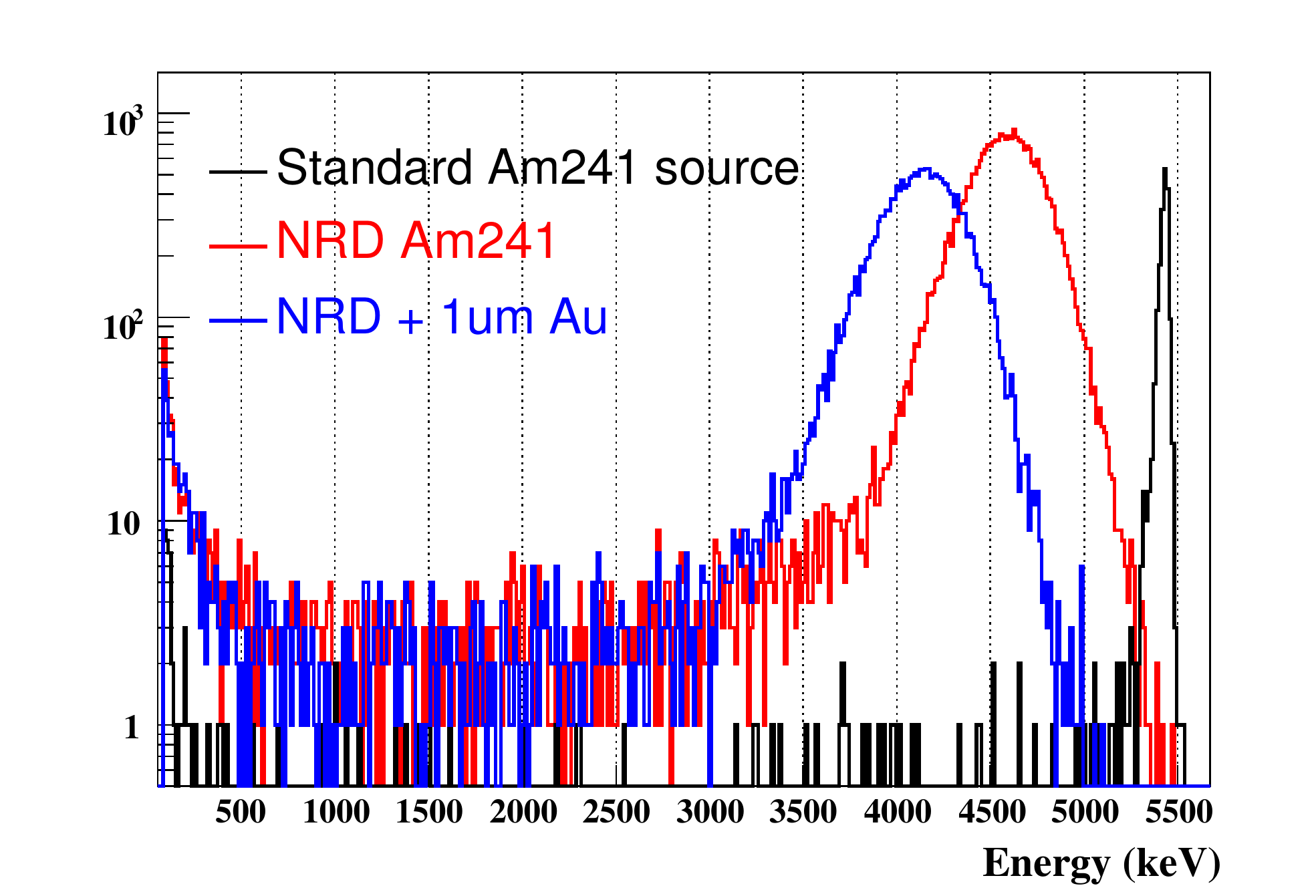}
  \caption{\it Measured alpha spectra with (blue) and without (red)
    the 1 $\mu$m gold foil. The spectrum from a standard $^{241}$Am  
    $\alpha$ source (5.5MeV)
    (black) is overlaid for reference.}
  \label{fig:nrd_attenuated}
\end{figure}

$^{241}$Am discs from NRD Inc., 5-mm in diameter, were procured with 4.5 MeV 
alpha energy as 
a key specification (custom energy was achieved by varying the thickness 
of the electrodeposited gold
coating).  The activity of 
$^{241}$Am is approximately 28 $\mu$Ci, and is deposited on 
one side of the disc only. 
Measurements of the emitting alpha energy were performed at Caltech 
in a vacuum 
chamber with 
a Si detector. The raw energy spectrum 
from the $^{241}$Am source is shown in Fig.~\ref{fig:nrd_attenuated}. 
Also overlaid is the energy spectrum from a standard $^{241}$Am 
source~\footnote{A typical
alpha calibration source has a thin front window of 100$\mu$g/cm$^2$, which
attenuates the 5.5 MeV alpha energy only by about 22 keV.}.
It was discovered that although the alpha energy for the NRD sources is 
peaked around 4.6 MeV, the distribution is rather broad,
all the way up to 5.5 MeV. 
The same measurements were performed 
on multiple discs and the results were consistent. 
In order to further reduce the alpha energy, 1~$\mu$m thick gold 
foil was purchased from Alfa Aesar and attached to the front surface 
of a NRD source. The attenuated energy spectrum is overlaid in Fig.~\ref{fig:nrd_attenuated}.
Compared to that without the gold foil, the entire energy spectrum 
was shifted by about 0.5 MeV, as expected.
Out of about 20000 total alphas, no events were
observed beyond 5.11 MeV threshold. 
\subsection{Expected neutron rate and energy spectrum}
\begin{figure}[!htbp]
\centering
\subfigure[\it Energy spectrum]
{    \label{fig:neutronEnergy}                                                         
    \includegraphics[width=0.45\textwidth]{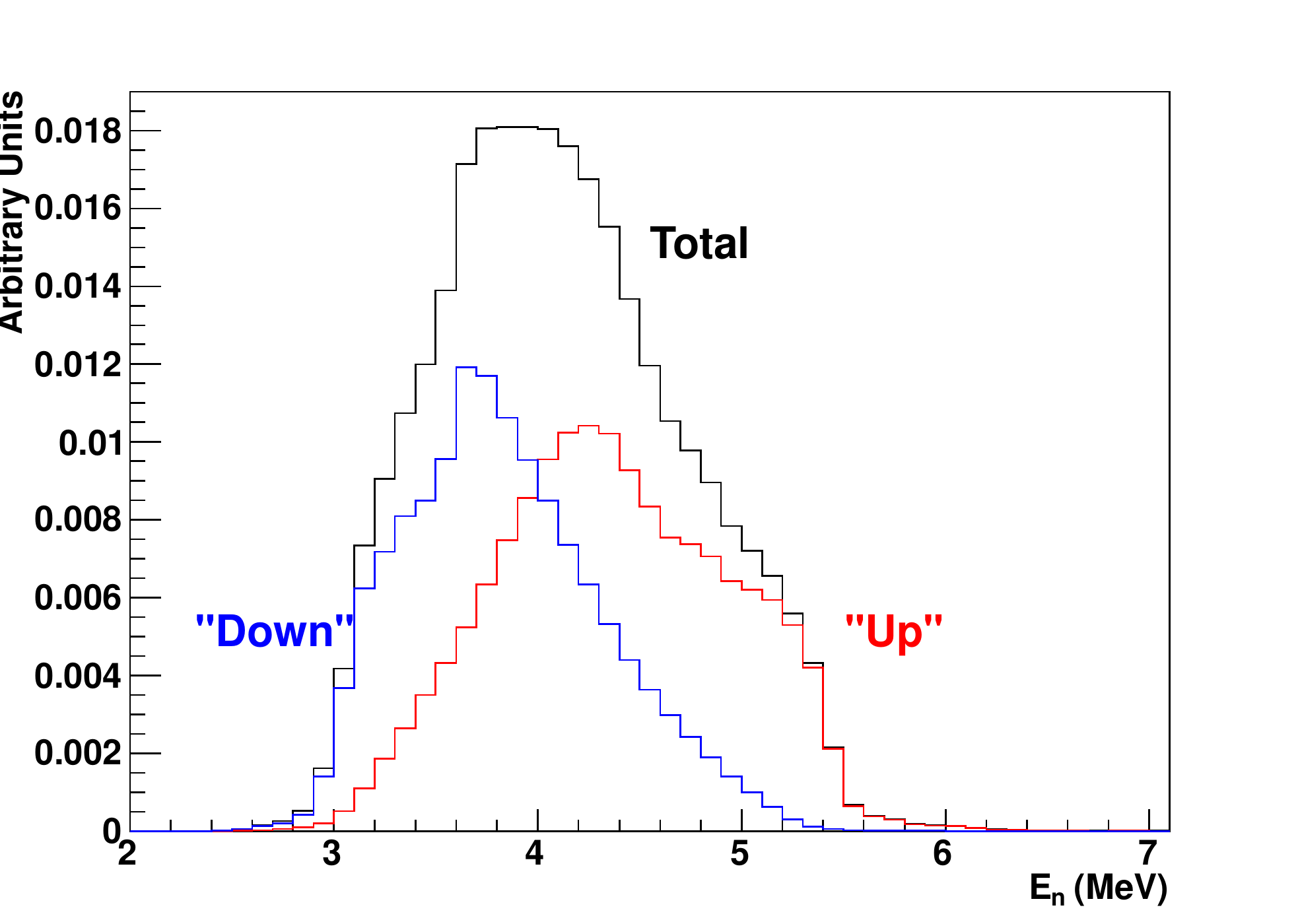}                                 
}
%\hspace{1cm}                                                                   
\subfigure[\it Energy-angle correlation] % caption for subfigure b                             
{                                                                               
  \label{fig:energy_angle}                                                          
  \includegraphics[width=0.45\textwidth]{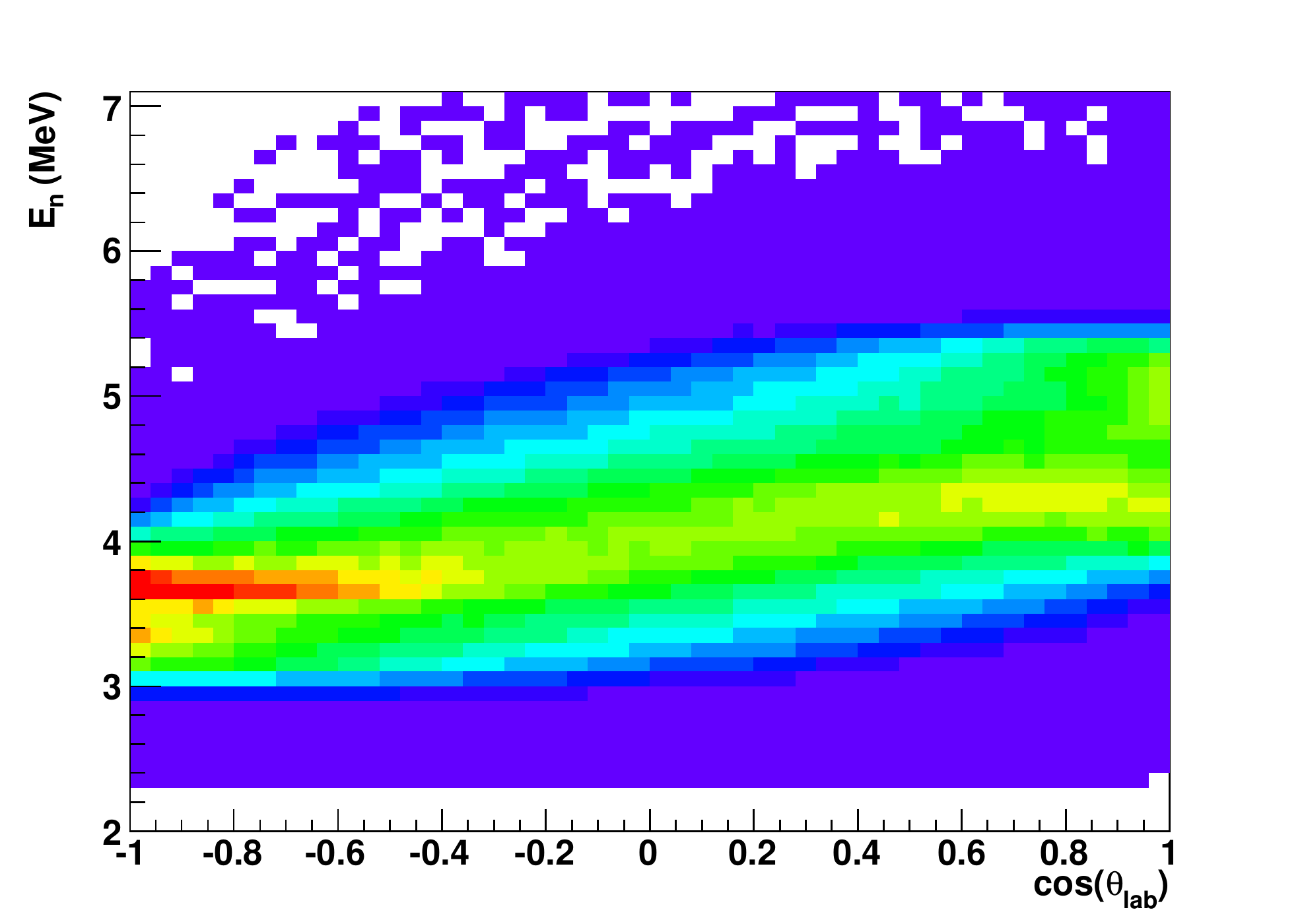}                                  
}
\caption{\it                                                                    
  a) Simulated neutron energy spectrum from the $^{241}$Am-$^{13}$C source: black=total, 
  red=heading out from the active side of $^{241}$Am (``up''), blue=heading
  opposite to the active side of $^{241}$Am (``down''). The fraction of neutrons 
  with energy $>$4.4 MeV is 33\%, 50\%, and 16\% for ``total'', ``up'', 
  and ``down'' distributions, respectively. 
  b) Simulated neutron energy vs. angle distribution. See text for details.
}
\label{fig:neutron_energy_energy} % caption for the whole figure  
\end{figure}

A GEANT4 program was developed to calculate the neutron rate from the source. 
To simulate the one-sided NRD source, the alphas were generated 
in random directions in the active hemisphere with 
energy sampled from the spectrum in Fig.~\ref{fig:nrd_attenuated} (the red histogram). 
The energy loss of the alphas in 1~$\mu$m gold and $^{13}$C was simulated by 
GEANT4. 
GEANT4 tracks alphas until they are stopped. 
When an alpha enters $^{13}$C, for each step $i$
along the track, one computes a step weight (which gets summed at 
the end of the event)
\begin{equation}
\label{eq:weight0}
\mathrm{weight_{i}} = \sigma(E_{\alpha,i})\times d_i\,,
\end{equation}
where $E_{\alpha,i}$ is the mean alpha energy in this step, $d_i$ is the 
step length and $\sigma(E_{\alpha,i})$ is the $^{13}$C($\alpha$,n) 
reaction cross section. The total neutron rate can then be calculated as 
\begin{equation}
\label{eq:rate}
  R_n = R_{\alpha} \times \displaystyle\sum_{i} \mathrm{weight_i}
  \times \displaystyle\frac{\rho_{^{13}C} \cal{N}_A}{13}\,
\end{equation}
where $R_{\alpha}$ is the alpha emission rate, $\rho_{^{13}C}$ is the density 
of $^{13}$C, and $\cal{N}_A$
is the Avogadro's number.
The outgoing neutron energy spectrum was obtained by generating a 
neutron in random direction (but not tracked) with respect to the alpha momentum
in each tracking step in $^{13}$C, with its energy calculated 
based on 2-body elastic scattering kinematics. Each neutron was assigned 
a step weight as in Eqn.~\ref{eq:weight0}.
The resulting neutron energy spectrum is shown in 
Fig.~\ref{fig:neutronEnergy}.
One sees that on average the neutron energy is 4 MeV, with a 
tail extending to 6.5 
MeV or so. The neutron energy-angle (where $\theta$ is the angle 
relative to the normal of the active alpha surface) correlated distribution is shown 
in Fig.~\ref{fig:energy_angle}. Clearly, neutrons heading 
opposite to the ``active'' side of the alpha source have significantly 
lower energy. 

High energy neutrons could inelastically scatter with materials surrounding 
and inside the detector, e.g. stainless steel, $^{12}$C, etc, and produce
gamma rays along the way, introducing correlated background 
when being detected in delayed coincidence with the 
final SS-capture gammas. To reduce such a residual background, we 
chose to use the source with the active side of the $^{241}$Am facing up to
reduce the energy of the downward going neutrons. 

\section{Mechanical Design and Fabrication}
\label{sec:fab}
Pure $^{13}$C is immediately available in powder form. To get a good neutron flux,
it is important to have the $^{241}$Am in close and uniform contact with $^{13}$C. 
Alpha leakage is a serious contamination,
so such a source should also be very safely sealed.

\begin{figure}[!htbp]
  \centering
  \includegraphics[width=3.5in]{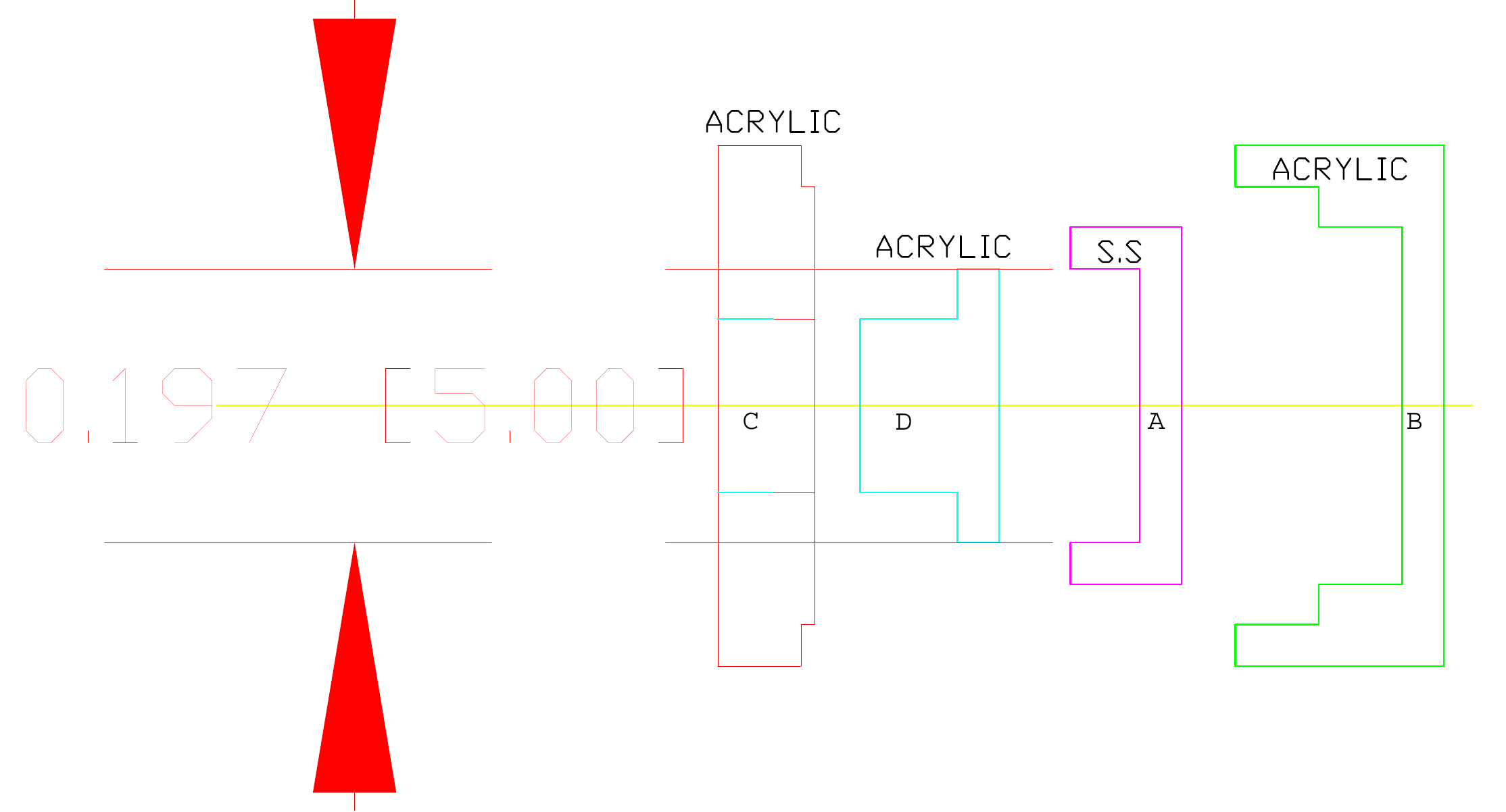}
  \caption{\it Mechanical drawing of the $^{241}$Am-$^{13}$C source assembly. See text for 
  more details.}
  \label{fig:enclosure}
\end{figure}

The mechanical design 
\footnote{A switchable source design \cite{switchable_source} was considered to further
reduce background, but it was given up due to the complexity in automating the switch.}
of the $^{241}$Am-$^{13}$C source is shown in 
Fig.~\ref{fig:enclosure}. 
A stainless cup with 5 mm inner diameter is the holder for $^{13}$C, which is 
pressed into a solid form. $^{241}$Am source, Au foil, and $^{13}$C
are sandwiched tightly. The stainless cup will be enclosed in an acrylic 
enclosure made out of three pieces: the bottom cup which houses the stainless 
cup; a plunger that presses $^{241}$Am, the Au foil, and $^{13}$C together,
and the top flange that seals the entire assembly. The seals between 
the acrylic pieces are made using the Weld-on 3 (volatile, water thin) 
acrylic cement \cite{weld_on}. Such a design went through a standard test procedure 
and obtained the State of California Certificate of 
Sealed Sources, which was a key requirement to transport these 
sources to Daya Bay. 

\section{Quality controls}
\label{sec:meas}
\begin{figure}[!htbp]
\centering
\subfigure[\it Top view] % caption for subfigure a
{
    \label{fig:topview}
    \includegraphics[width=0.45\textwidth]{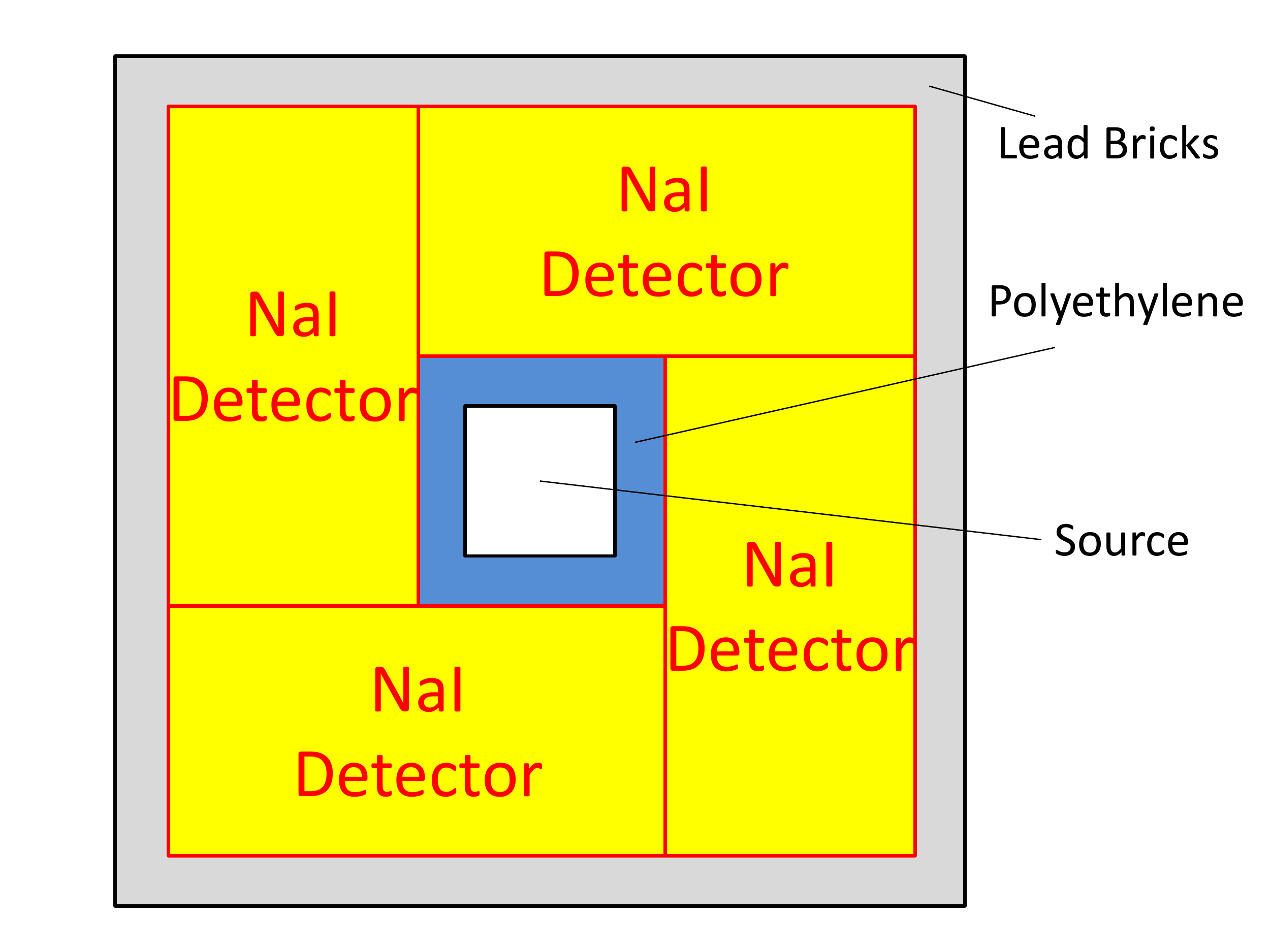}
}
%\hspace{1cm}
\subfigure[\it Side view] % caption for subfigure b
{
  \label{fig:sideview}
  \includegraphics[width=0.45\textwidth]{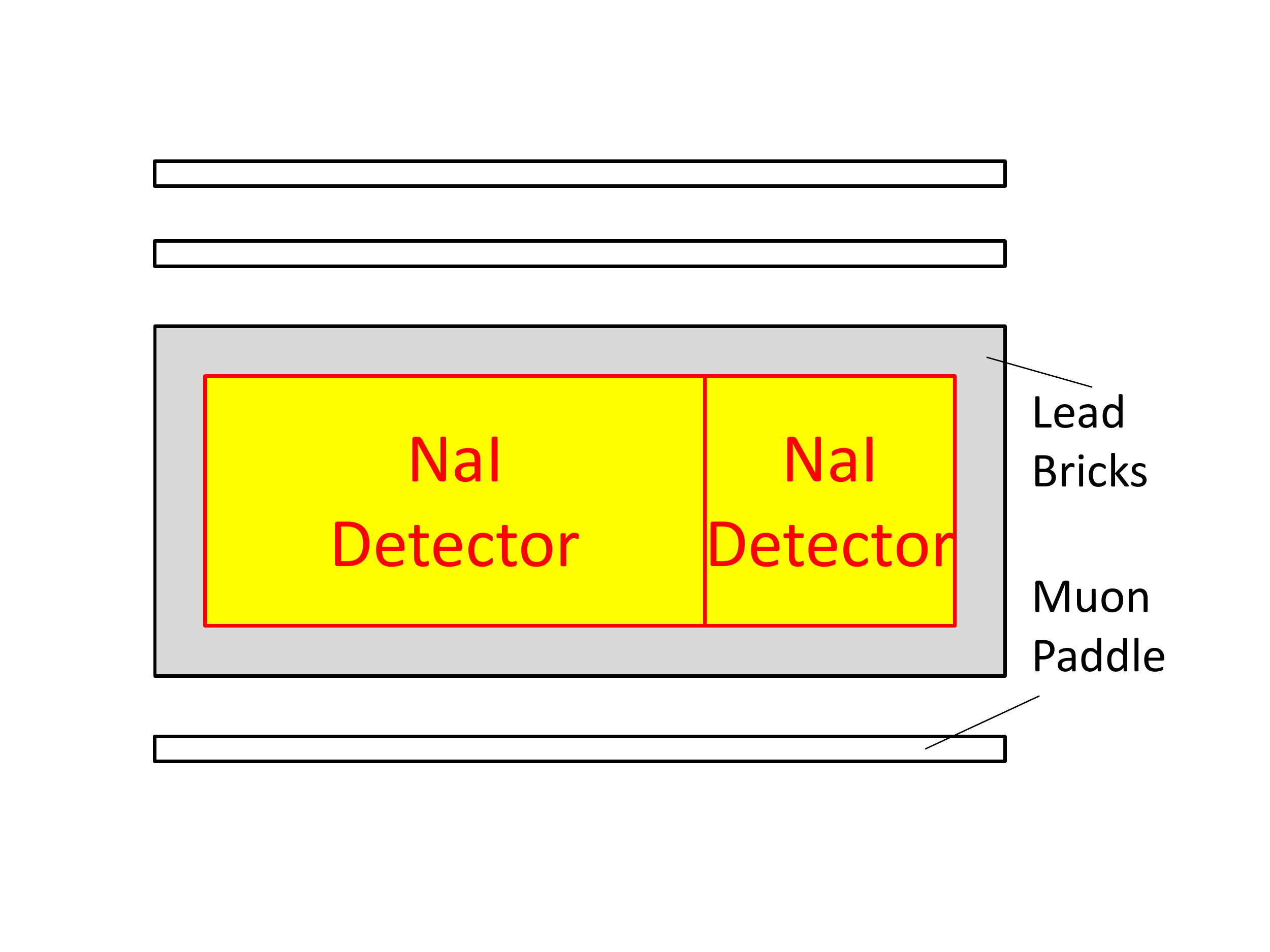}
}
\caption{\it 
  Schematic views of the neutron detector setup.
}
\label{fig:ndet_schematic} % caption for the whole figure
\end{figure}

Twenty-eight $^{241}$Am-$^{13}$C sources were fabricated at Caltech, 
twenty-four of which were transported to 
Daya Bay after a quality control (QC) process. During QC,
we set up a neutron detector in a 
low background environment, and measured the neutron rates
from these sources. 
The detector assembly consists of an array of 4  
NaI detectors (15$\times$15$\times$30 cm), and neutrons are detected via 
\begin{equation}
  {}^{127}I + n \rightarrow {}^{128}I + \gamma \rm (\sim 6 MeV)\,.
\end{equation}
Schematic views of the setup are shown in Fig.~\ref{fig:ndet_schematic}.
Due to the relatively low neutron emission rate in our $^{241}$Am-$^{13}$C sources, 
external background had to be significantly reduced 
to achieve a tolerable signal to background ratio.
We placed a detector into a subbasement lab on Caltech campus, with 
an overburden of about 6 meters consisting of 1 meter of 
concrete and 5 meters of dirt to shield against cosmogenic neutron 
background. The assembly was shielded against ambient 
gamma rays by lead bricks of about 2 inch (5.08 cm) thick. 
Two scintillator paddles 
($\sim$ 60 cm $\times$ 60 cm $\times$ 2.5 cm)
were placed on the top and one on the bottom of the assembly to
serve as a muon veto with an efficiency measured to be $\sim$ 95\%.
The $^{241}$Am-$^{13}$C neutron source to be assayed was placed at 
the center of the detector assembly to maximize the acceptance. 
The threshold of each NaI was set at $\sim$3 MeV. Any over-threshold 
hit in a NaI generated a first level trigger which, in combination with
the veto signals from the muon paddles (30$\mu$s window), formed the main 
trigger to read out the ADCs/TDCs. 
The residual background neutron-like rate, 
summed over all four NaI detectors, is about 1.4 (0.25) Hz without (with)
the muon veto enforced.

\begin{figure}[!htbp]
  \centering
  \includegraphics[width=4in]{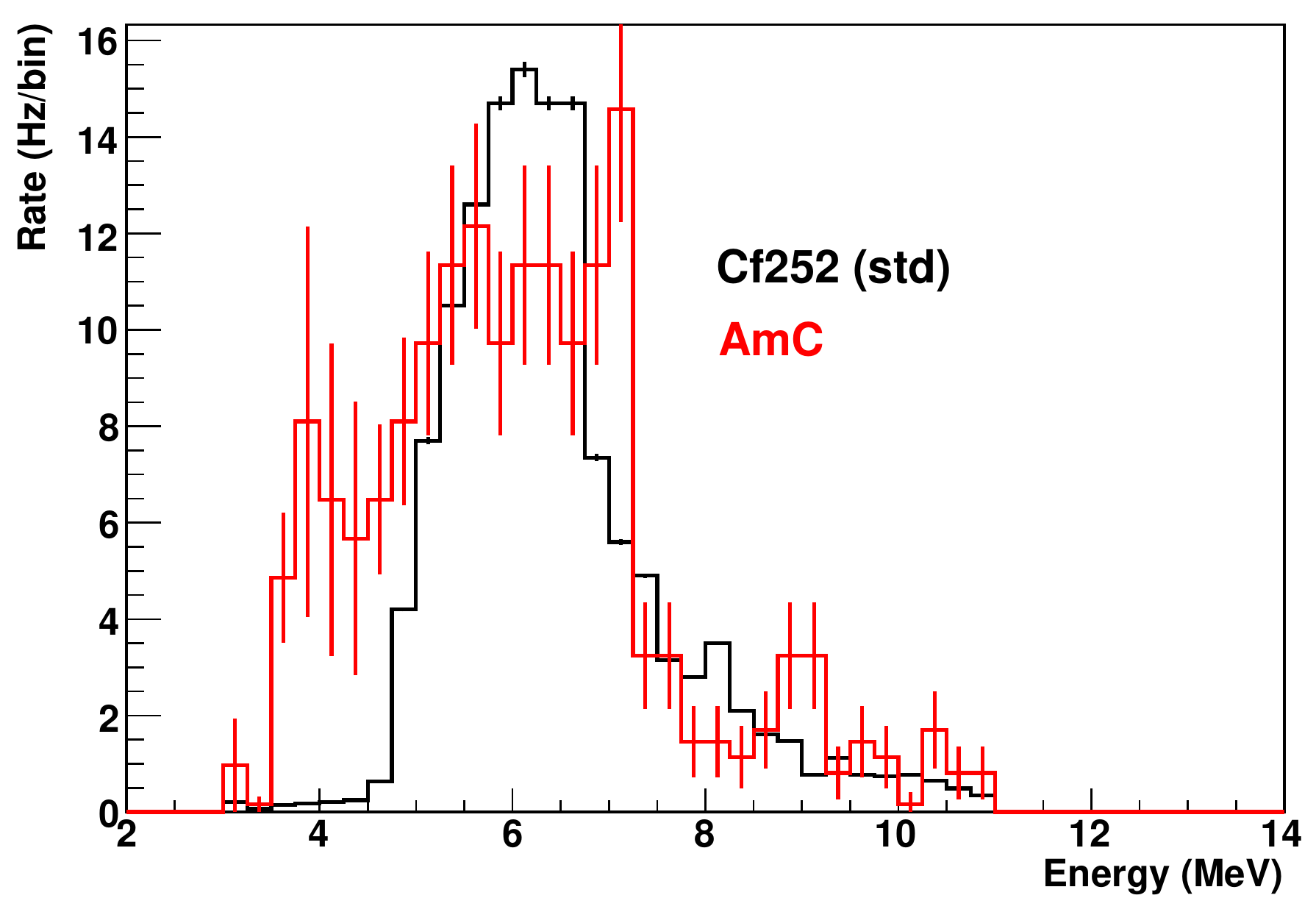}
  \caption{\it Detected energy spectra for neutron sources 
    with background subtracted.
    Black: spectrum from standard $^{252}$Cf, 
    Red: a typical $^{241}$Am-$^{13}$C source, scaled 
    for visual clarity.}
  \label{fig:n_det_demo}
\end{figure}

\begin{figure}[!htbp]
  \centering
  \includegraphics[width=4in]{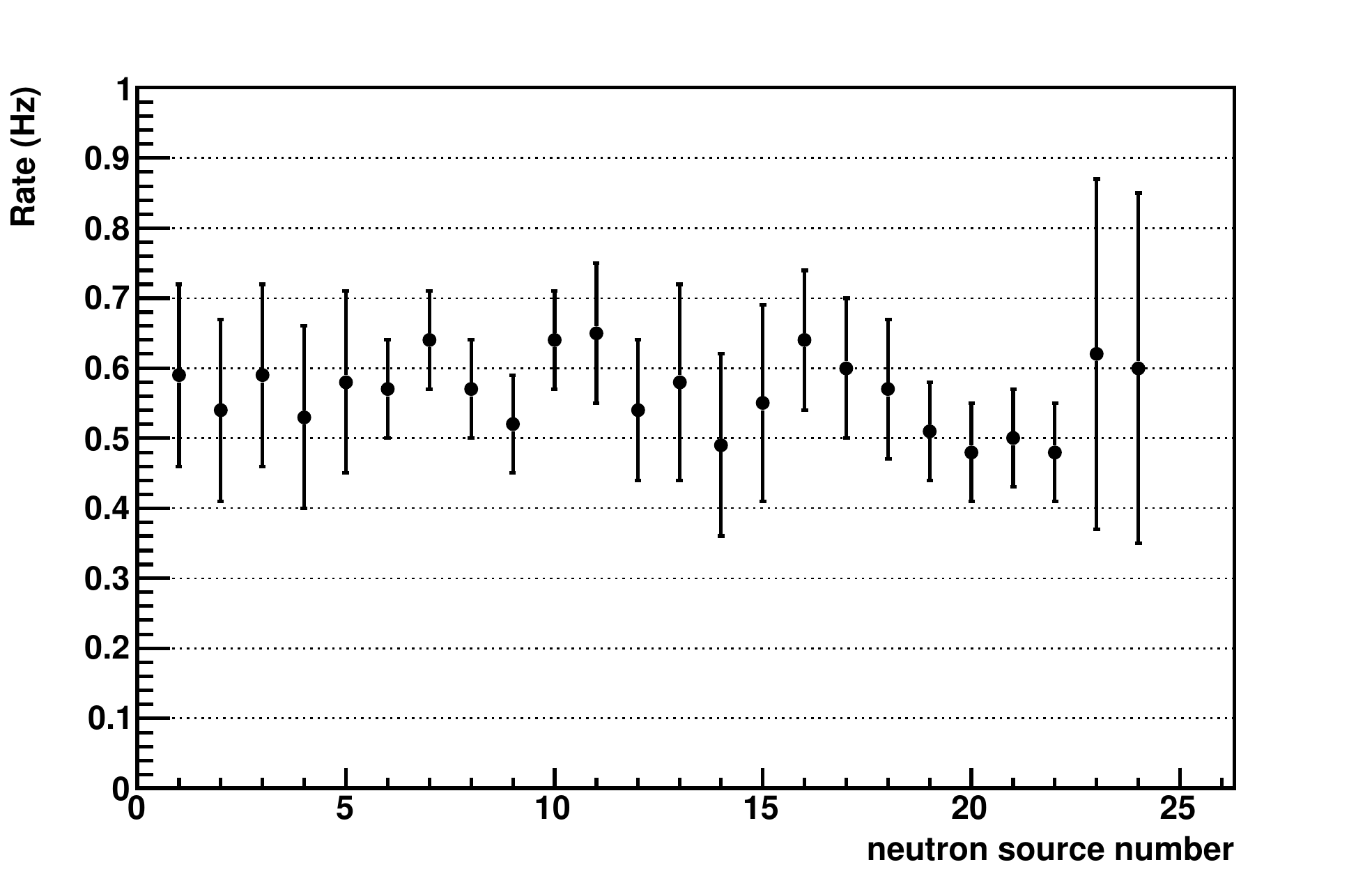}
  \caption{\it Summary of the rate of the neutron sources measured on the bench. 
  Only statistical errors are shown.
  The larger error bars of the last two points were due to an increase of background.}
  \label{fig:nrate}
\end{figure}

The neutron detection efficiency was calibrated with a 
standard 2.7 kHz $^{252}$Cf neutron source. 
The energy spectrum of the capture gammas
is shown in Fig.~\ref{fig:n_det_demo}. 
One clearly observes the iodine capture gamma ``bump''. 
When setting an energy cut between 4.5 and 8 MeV, 
the neutron detection efficiency is calibrated to be $5\pm1\%$, where the uncertainty
is dominated by the absolute neutron rate.
For our weak $^{241}$Am-$^{13}$C sources, a measurement cycle consists of a 
24-hour overnight background counting run, followed by a 24-hour overnight ``signal'' 
run. Statistical subtraction was made to extract the neutron source
signals.
The shape of the extracted neutron signals is roughly consistent with 
that from the calibrated sources in Fig.~\ref{fig:n_det_demo}.
The measured neutron rates of all 24 neutron sources are summarized 
in Fig.~\ref{fig:nrate}. 
The average neutron rate per source is 0.56 Hz, with a maximum variation of $\pm$15\% 
from source to source and an overall systematic 
uncertainty of 20\% due to detection efficiency. This clearly satisfies our low rate 
requirement highlighted in Sec.~\ref{sec:requirements}.

\section{Performance of the neutron sources at Daya Bay}
\label{sec:dayaanalysis}
Neutron sources were shipped to Daya Bay and installed in the ACUs during the 
detector assembly (detailed in \cite{ref:ACU}). 
To achieve a minimum background, at the parking location inside the 
ACU the neutron source is surrounded by a borated polyethylene cylinder (BPE) 
with 5 inch height and 2.25 inch wall thickness. Below the cylinder, 
there is another BPE disk with 3.25 inch in diameter and 2.5 inch in 
thickness acting as further shield.

\begin{figure}[!htbp]
\centering
\subfigure[\it Prompt energy spectrum] % caption for subfigure a                              
{                                                                               
    \label{fig:Ep}                                                         
    \includegraphics[width=0.4\textwidth]{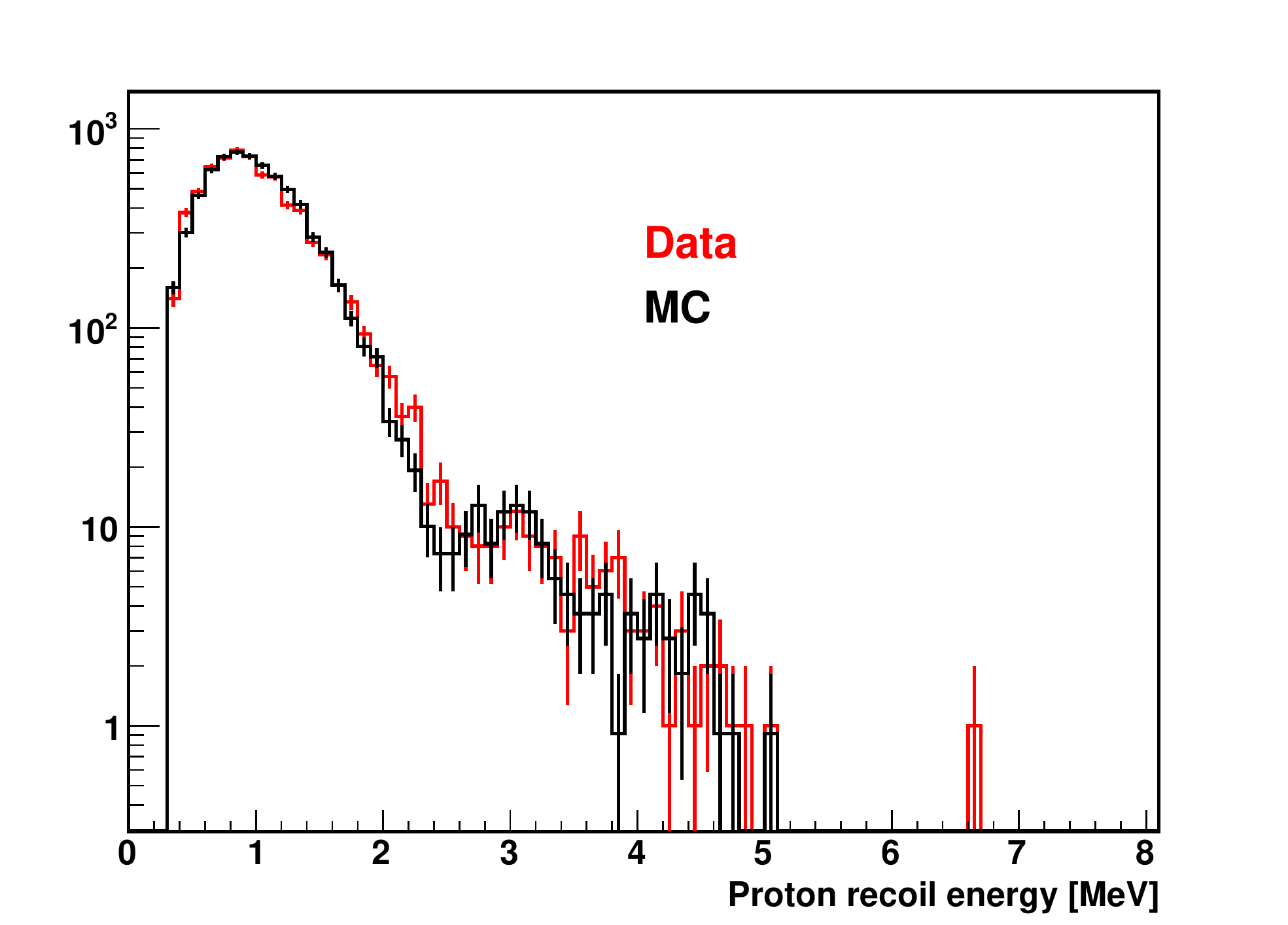}                         
}
%\hspace{1cm}                                                                   
\subfigure[\it Delayed energy spectrum]
{                                                                               
  \label{fig:Ed}                                                          
  \includegraphics[width=0.4\textwidth]{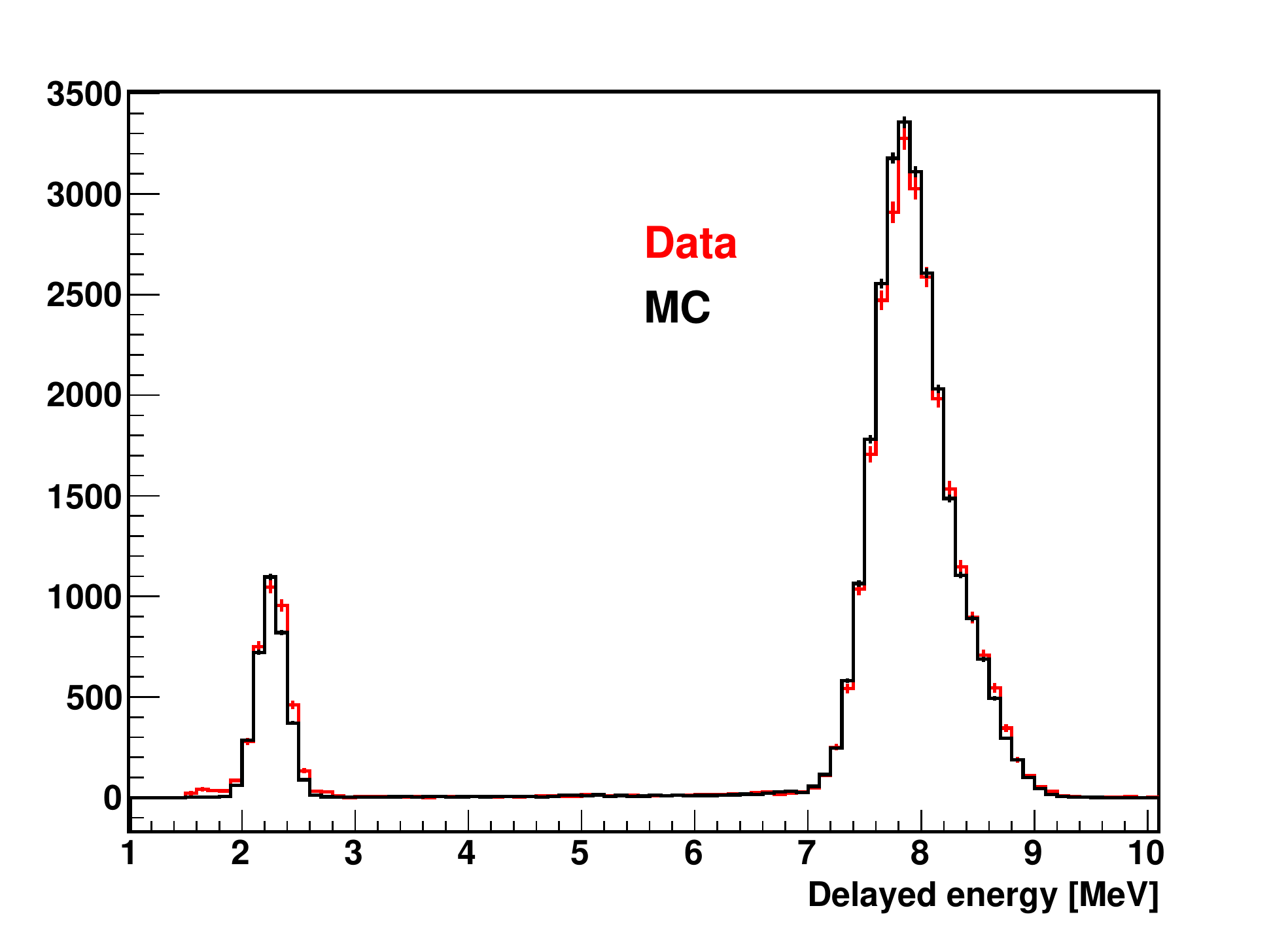}
}
\subfigure[\it Prompt-delay time separation] % caption for subfigure b
{
  \label{fig:dt}                                                                
  \includegraphics[width=0.4\textwidth]{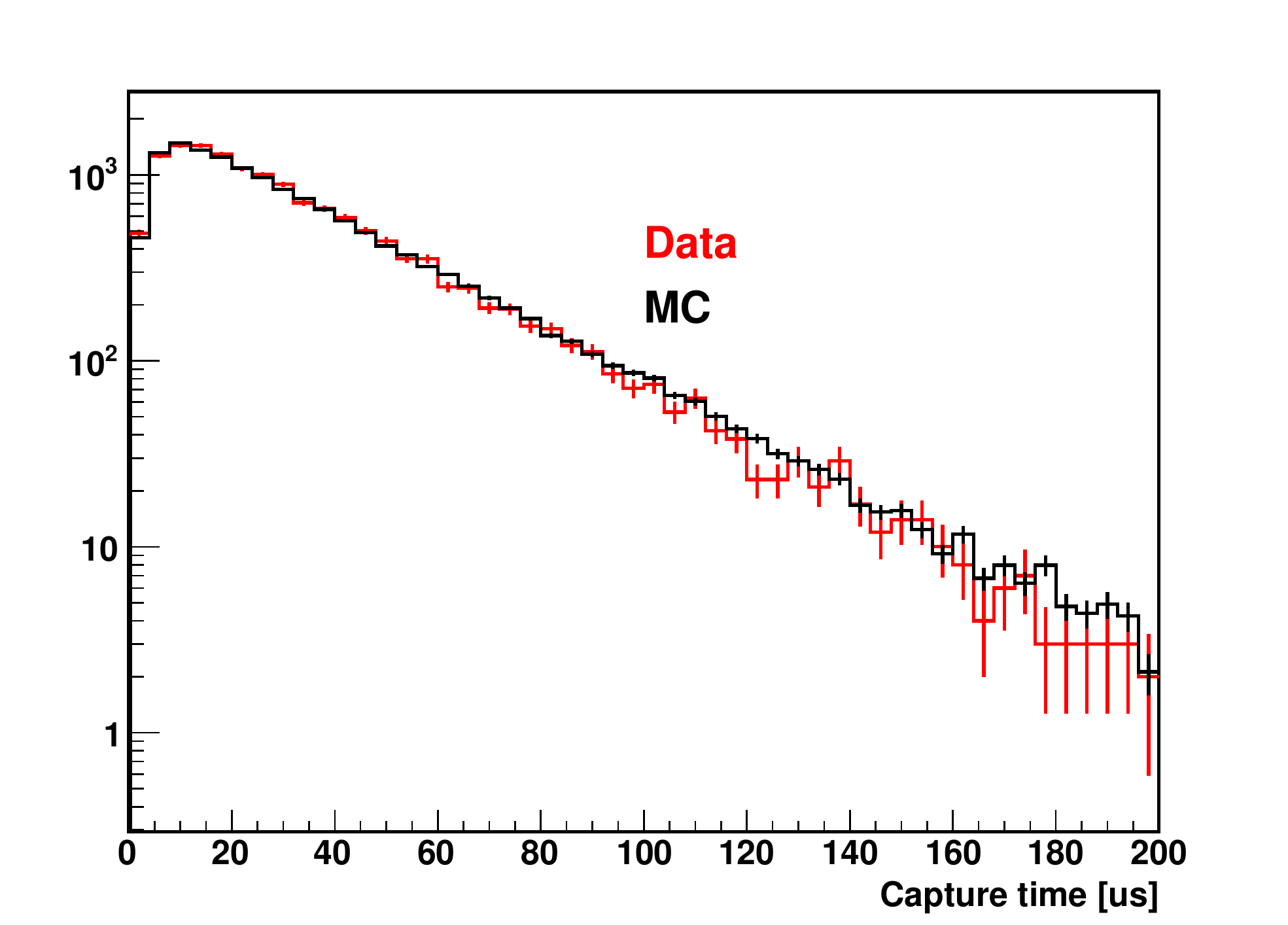}                                
}
\caption{\it                                                                    
  Reconstructed prompt (a) and delayed (b) energy spectra, as well as the time separation
  between the two (c). The data and MC simulation are normalized to equal area and overlaid.                                  
}
\label{fig:AmC_on-site} % caption for the whole figure                       
\end{figure}

Neutron sources are used extensively in calibrating the detector 
response, which has been reported in \cite{DYBNIM12,DYBCPC}. 
We limit the discussions 
here to some basic performance of these sources. 
When the sources are deployed into the detector,
the proton recoil and the 
following capture 
gammas form time-correlated pairs, similar to IBD signals.
The average neutron rate 
as measured in the detector is 0.7 Hz, consistent with the QC measurements at Caltech.
In Fig.~\ref{fig:AmC_on-site}, the prompt-delayed energy spectrum, as well as 
the time separation in between for an $^{241}$Am-$^{13}$C source deployed in an AD is shown, overlaid 
with the expected distribution from MC. All distributions 
agree well with MC expectations. We observe no evidence 
of the 6.1 MeV gamma emission, confirming that the key design specification is 
met. The residual background introduced by these neutrons is quite 
small~\cite{DYB12}. The details of the evaluation will be discussed in 
a separate article~\cite{ref:AmC}. 

\section{Summary}
We have discussed the design and construction of the special compact 
and low rate neutron sources in the Daya Bay experiment. 
The performance of the low background neutron sources has been proven
to satisfy the design specifications at Daya Bay. Such sources could be
used in other experiments requiring low backgrounds.

\label{sec:summary}

\section{Acknowledgments}
This work was done with support from the US DoE, 
Office of Science, High Energy Physics, the US National Science Foundation, 
the Natural Science Foundation of China Grants 11175116, the Chinese 
MOST grant 2013CB834306, and Shanghai Laboratory for Particle Physics and Cosmology at 
the Shanghai Jiao Tong University.
This work is supported in part by the CAS Center for 
Excellence in Particle Physics (CCEPP).

We gratefully thank Dick Hahn from BNL
for his critical guidance in making these 
sources. We also thank David Jaffe from BNL
for his intellectual input. The technical support from 
the Kellogg technical team Ray Cortez and Jim Pendlay,
particularly Ray's design work, is truly indispensable. We also appreciate
the safety guidance from the Caltech radiation safety officers Haick 
Issaian and Andre Jefferson, and on-site logistical support 
from Xiaonan Li of IHEP. 

%% The Appendices part is started with the command \appendix;
%% appendix sections are then done as normal sections
%% \appendix

%% \section{}
%% \label{}

%% References
%%
%% Following citation commands can be used in the body text:
%% Usage of \cite is as follows:
%%   \cite{key}          ==>>  [#]
%%   \cite[chap. 2]{key} ==>>  [#, chap. 2]
%%   \citet{key}         ==>>  Author [#]

%% References with bibTeX database:

\bibliographystyle{model1-num-names}
\bibliography{<your-bib-database>}

%% Authors are advised to submit their bibtex database files. They are
%% requested to list a bibtex style file in the manuscript if they do
%% not want to use model1-num-names.bst.

%% References without bibTeX database:

\end{document}